\documentclass[12pt]{article}
\usepackage{amsmath}
\usepackage{graphics,graphicx,xcolor}

\def\be{\begin{equation}}
\def\ee{\end{equation}}
\def\arr{\begin{array}{rll}}
\def\ea{\end{array}}
\def\bea{\begin{eqnarray}}
\def\eea{\end{eqnarray}}
\def\N2{$N{=}2$}

\def\>{\rangle}
\def\<{\langle}
\def\+{\dagger}
\def\={\ =\ }

\def\bal{\begin{aligned}}
\def\eal{\end{aligned}}
\begin{document}
\begin{titlepage}
\setcounter{page}{0}
\begin{center}
{\LARGE\bf  Remarks on $\mathcal{N}=1$ supersymmetric }\\
\vskip 0.4cm
{\LARGE\bf extension of the Euler top }\\
\vskip 1.5cm
\textrm{\Large Anton Galajinsky \ }
\vskip 0.7cm
{\it
Tomsk State University of Control Systems and Radioelectronics, 634050 Tomsk, Russia} \\

\vskip 0.2cm
{e-mail: a.galajinsky@tusur.ru}
\vskip 0.5cm
\end{center}

\begin{abstract} \noindent
A natural $\mathcal{N}=1$ supersymmetric extension of the Euler top, which introduces exactly one fermionic counterpart for each bosonic degree of freedom, is considered.
The equations of motion, their symmetries and integrals of motion are given. It is demonstrated that, although in general the system lacks the integrability property, it admits an interesting integrable reduction, for which all fermions are proportional to one and the same Grassmann--odd number - a value of the conserved supercharge. A generalisation involving an arbitrary three--dimensional real Lie algebra is proposed.
\end{abstract}

\vspace{0.5cm}

\end{titlepage}
\renewcommand{\thefootnote}{\arabic{footnote}}
\setcounter{footnote}0

\noindent
{\bf 1. Introduction}\\

\noindent
Supersymmetric extensions of classical mechanics models were extensively studied in the past few decades while they were being encountered within string theory (super 0--branes) and black hole physics (microscopic description of near horizon geometries).
An interrelation between supersymmetry and integrability attracted less attention, though. It is widely believed that a supersymmetric extension of an integrable system should automatically result in a new integrable model. If this were the case, a supersymmetrisation procedure would provide an efficient means of generating novel integrable systems.

Technically speaking, the construction of an $\mathcal{N}=1$ supersymmetric extension of a mechanical system with the Hamiltonian $\mathcal{H}_0$ amounts to introducing fermionic partners for bosonic degrees of freedom and constructing a {\it single} real supersymmetry charge $Q$, which via the Poisson bracket $\{Q,Q \}=-2{\rm i} \mathcal{H}$ produces the super--extended Hamiltonian $\mathcal{H}$. The latter governs the dynamics of the enlarged system and reduces to $\mathcal{H}_0$  in the limit of vanishing fermions. Because in general one has several fermionic degrees of freedom and only one conserved supercharge, integrability of the ensuing system in the fermionic sector is not a priori guaranteed.

Aiming at a better understanding of the interplay between supersymmetry and integrability, in this work we study a natural $\mathcal{N}=1$ supersymmetric extension of the Euler top, in which bosons and fermions are present in equal numbers.  Besides purely academic interest, supersymmetric models of such a kind might be relevant for describing superconformal mechanics with spin degrees of freedom (see e.g. \cite{FIL1}--\cite{AG1} and references therein). Although the spin sector of many of them is described by the Euler top, a supersymmetric extension of the internal sector is usually ignored.

The work is organised as follows.

In the next section, we construct an $\mathcal{N}=1$ supersymmetric extension of the Euler top within the Hamiltonian framework by introducing exactly one fermionic counterpart for each bosonic degree of freedom. A simple supersymetry generator linear in the fermionic variables suffices to produces a reasonable super--extended Hamiltonian. The equations of motion, their symmetries and integrals of motion are given in explicit form.

In Sect. 3, Lax matrices are constructed which reproduce the equations of motion and first integrals obtained in Sect. 2.

The issue of integrability is discussed in Sect. 4. Because some integrals of motion are quadratic in fermions, one faces the difficulty in algebraically resolving them as there does not exist a division by a Grassmann--odd number.
Representing a general solution as a power series in Grassmann--odd constants of integration, one can turn the original supersymmetric equations of motion into a system of {\it bosonic} first order ordinary differential equations, which can be analysed by conventional means. It is shown that the system lacks the integrability property. Yet, an interesting integrable reduction can be constructed which corresponds to a particular solution to the original supersymmetric equations of motion, for which all fermions are proportional to one and the same Grassmann--odd number - a value of the conserved supercharge.

The Hamiltonian description of the Euler top relies upon $su(2)$ algebra, which is a particular instance in the Bianchi classification of three--dimensional real Lie algebras.
Sect. 5 generalises the analysis in Sect. 2 to incorporate arbitrary three--dimensional real Lie algebra. Both the equations of motion and conserved (super)charges are given.

We summarise our results and discuss possible further developments in the concluding Sect. 6.

Throughout the paper, summation over repeated indices is understood unless otherwise is stated explicitly.

\vspace{0.5cm}

\noindent
{\bf 2. $\mathcal{N}=1$ supersymmetric extension of the Euler top}\\

\noindent
The Euler top describes a rigid body freely rotating around its center of mass. Within the Hamiltonian framework, it is represented by the angular velocity vector (measured in a rotating frame) $J_i$, $i=1,2,3$, which obeys the (degenerate) Poisson bracket
\be\label{alg}
\{J_i,J_j \}=\epsilon_{ijk} J_k,
\ee
$\epsilon_{ijk}$ being the Levi--Civita symbol with $\epsilon_{123}=1$. The Hamiltonian is equal to the kinetic energy
\be\label{Ham}
H=\frac 12 a_i^2 J_i^2,
\ee
where $a_i$ are real constants related to the moments of inertia $I_i=\frac{1}{a_i^2}$. The equations of motion read
\be\label{eom}
{\dot{J}}_1=\left(a_2^2-a_3^2 \right) J_2 J_3, \qquad {\dot{J}}_2=-\left(a_1^2-a_3^2 \right) J_1 J_3, \qquad {\dot{J}}_3=\left(a_1^2-a_2^2 \right) J_1 J_2.
\ee

Regarding (\ref{alg}), (\ref{Ham}) as a dynamical system associated with $su(2)$ Lie algebra, one
concludes that the Casimir invariant
\be\label{CI}
\mathcal{I}=J_i J_i
\ee
is a constant of the motion. $\mathcal{I}$ and $H$ can then be used to express $J_1$ and $J_2$ in terms of $J_3$, while Eqs. (\ref{eom}) yield an elliptic integral which links $J_3$ to a temporal variable $t$.

The case $a_1=a_2=a_3$ describes a spherical Euler top
\be\label{sol1}
J_i=c_i,
\ee
where $c_i$ are real constants of integration. If two moments of inertia are equal to each other, say $a_2=a_3$, one has a symmetric Euler top
\be\label{sEt}
J_1=c_1, \qquad J_2=c_2 \cos{\left(\left(a_1^2-a_2^2\right)c_1 t+c_3 \right)}, \qquad J_3=c_2 \sin{\left(\left(a_1^2-a_2^2\right)c_1 t+c_3 \right)},
\ee
where $c_1$, $c_2$, and $c_3$ are constants of integration.

In order to construct an $\mathcal{N}=1$ supersymmetric extension of (\ref{Ham}), for each component $J_i$ one introduces a real fermionic partner $\theta_i$, obeying the Poisson brackets
\be
\{\theta_i,\theta_j \}=-{\rm i} \delta_{ij}, \qquad \{\theta_i,J_j \}=0,
\ee
and then builds the supersymmetry charge
\be\label{sc}
Q=a_i \theta_i J_i,  \qquad \{Q,Q \}=-2 {\rm i} \mathcal{H}.
\ee
The super--extended Hamiltonian $\mathcal{H}$ in (\ref{sc}) reads
\be\label{SHam}
\mathcal{H}=\frac 12 a_i^2 J_i^2 +\frac{\rm i}{2} \epsilon_{ijk} a_i \theta_i a_j \theta_j J_k,
\ee
which gives rise to the equations of motion
\be\label{EOM}
{\dot J}_i=\epsilon_{ijk} a_j^2 J_j J_k-{\rm i} a_{\hat i} \theta_{\hat i} Q, \qquad {\dot\theta}_i=a_{\hat i} \epsilon_{{\hat i} jk} a_j \theta_j J_k,
\ee
where the identity $\epsilon_{ijk} \epsilon_{plk}=\delta_{ip} \delta_{jl}-\delta_{il} \delta_{jp}$ was used.
In Eqs. (\ref{EOM}) and in the text below no summation over repeated indices carrying a hat symbol is understood. Note that $J_i$ and $\theta_i$ take values in the even and odd parts of the infinite--dimensional Grassmann algebra, respectively. The component form of (\ref{EOM}) reads
\begin{align}\label{seom}
&
{\dot{J}}_1=\left(a_2^2-a_3^2 \right) J_2 J_3-{\rm i} a_1 \theta_1 Q, && {\dot{J}}_2=-\left(a_1^2-a_3^2 \right) J_1 J_3-{\rm i} a_2 \theta_2 Q,
\nonumber\\[2pt]
&
{\dot{J}}_3=\left(a_1^2-a_2^2 \right) J_1 J_2-{\rm i} a_3 \theta_3 Q, && \dot\theta_1=a_1 a_2 \theta_2 J_3-a_1 a_3 \theta_3 J_2,
\nonumber\\[2pt]
&
\dot\theta_2=-a_1 a_2 \theta_1 J_3+a_2 a_3 \theta_3 J_1, && \dot\theta_3=a_1 a_3 \theta_1 J_2-a_2 a_3 \theta_2 J_1.
\end{align}

As is well known, within the Hamiltonian framework infinitesimal canonical symmetry transformations are generated by conserved charges themselves. Abbreviating
\be
\delta_\epsilon A={\rm i} \{ A,Q\} \epsilon, \qquad  \delta_\alpha A=\{ A,\mathcal{H}\} \alpha,
\ee
where $A$ is an arbitrary function on the phase space and $\epsilon$ and $\alpha$ are infinitesimal (super)translation parameters, one gets
\begin{align}\label{comp1}
&
\delta_\epsilon \theta_i=a_{\hat i} J_{\hat i} \epsilon, && \delta_\epsilon J_i={\rm i} \epsilon_{ijk} a_j \theta_j J_k \epsilon;
\nonumber\\[2pt]
&
\delta_\alpha \theta_i=a_{\hat i} \epsilon_{\hat i jk} a_j \theta_j J_k \alpha, && \delta_\alpha J_i=\epsilon_{ijk} a_j^2 J_j J_k \alpha -{\rm i} a_{\hat i} \theta_{\hat i} Q \alpha.
\end{align}
In this notation, the $d=1$, $\mathcal{N}=1$ supersymmetry algebra reads
$[\delta_{\epsilon_1},\delta_{\epsilon_2}]=\delta_\alpha$ with $\alpha=2{\rm i} \epsilon_1 \epsilon_2$. The transformations (\ref{comp1}) will prove helpful when constructing a Lax pair formulation for Eqs. (\ref{seom}) in Sect. 3.

Let us discuss integrals of motion of the supersymmetric extension (\ref{seom}).
As in the purely bosonic case, the $su(2)$ Casimir element $\mathcal{I}=J_i J_i$ commutes with $\mathcal{H}$. Because a square of a fermion is zero, the cubic combination
\be\label{Om}
\Omega=\frac{1}{3!}\epsilon_{ijk} \theta_i \theta_j \theta_k=\theta_1 \theta_2 \theta_3, \qquad \{\Omega,\Omega \}=0,
\ee
is conserved over time. The Poisson bracket of $Q$ and $\Omega$ then yields an extra constant of the motion
\be\label{La}
\Lambda={\rm i} \epsilon_{ijk} a_i J_i \theta_j \theta_k, \qquad \{Q,\Omega \}=-\frac 12 \Lambda, \qquad \{\Lambda,Q \}=0, \qquad \{\Lambda,\Omega \}=0,
\ee
which is quadratic in the fermionic variables. The identity
\be
Q \Lambda=4 {\rm i} \mathcal{H} \Omega,
\ee
implies that $\Omega$ is functionally dependent, though. In a similar fashion one can rule out $a_i^2 J_i^2 Q$, $a_i^2 J_i^2 \Lambda$, and $a_i^2 J_i^2 \Omega$ from the set of functionally independent integrals of motion.

Although the cubic integral $\Omega$ is not independent, it proves helpful in establishing the fact that $\mathcal{H}$, $\mathcal{I}$, $Q$, and $\Lambda$ comprise all the functionally independent integrals of motion of the dynamical system (\ref{seom}). Indeed, assume that the model admits one more real fermionic invariant $F=q_1 \theta_1+q_2 \theta_2+q_3 \theta_3+{\rm i} q_4 \theta_1 \theta_2 \theta_3$, where $q_1,\dots,q_4$ are bosonic functions of $J_i$ to be fixed from the equation $\{F,\mathcal{H}\}=0$. If $F$ were a first integral of (\ref{seom}), the product
\be
F \Lambda=2 {\rm i} \left( a_i q_i J_i \right)  \Omega
\ee
would be a conserved quantity as well. The last relation implies
\be
q_i=\frac{J_{\hat i}}{a_{\hat i}},
\ee
with $i=1,2,3$, and yields no restriction upon $q_4$. A direct inspection of $\{F,\mathcal{H}\}=0$ then leads to a contradiction. In a similar fashion, one can rule out an extra quadratic invariant $q_1 \theta_2 \theta_3+q_2 \theta_1 \theta_3+q_3 \theta_1 \theta_2$. It suffices to multiply the latter with $Q$ and take into account the fact that $\Omega$ is conserved.

\vspace{0.5cm}

\noindent
{\bf 3. A Lax pair formulation}\\

\noindent
It is interesting to compare the results in the preceding section with a Lax pair formulation. Starting with two anti--symmetric matrices
\be\label{LR}
L=\left(
\begin{array}{cccc}
0 & J_1 & J_2 \\
-J_1 &   0 & J_3 \\
-J_2 &   -J_3 & 0 \\
\end{array}
\right), \qquad
R=\left(
\begin{array}{cccc}
0 & \theta_1 & \theta_2 \\
-\theta_1 &   0 & \theta_3 \\
-\theta_2 &   -\theta_3 & 0 \\
\end{array}
\right),
\ee
and taking into account the supersymmetry transformations (\ref{comp1}) and the supersymmetry algebra $[\delta_{\epsilon_1},\delta_{\epsilon_2}]=\delta_\alpha$, one can represent the equations of motion (\ref{seom}) in the Lax form
\begin{align}\label{LP}
&
\dot L=[L,M], &&
M=\left(
\begin{array}{cccc}
0 & a_1^2 J_1+{\rm i} a_2 a_3 \theta_2 \theta_3  & a_2^2 J_2-{\rm i} a_1 a_3 \theta_1 \theta_3 \\
-a_1^2 J_1-{\rm i} a_2 a_3 \theta_2 \theta_3  &   0 & a_3^2 J_3+{\rm i} a_1 a_2 \theta_1 \theta_2  \\
-a_2^2 J_2+{\rm i} a_1 a_3 \theta_1 \theta_3  &   -a_3^2 J_3-{\rm i} a_1 a_2 \theta_1 \theta_2  & 0 \\
\end{array}
\right),
\nonumber\\[6pt]
&
\dot R=[R,S], && S=\left(
\begin{array}{cccc}
0 & -a_2 a_3 J_1 & -a_1 a_3 J_2 \\
a_2 a_3 J_1 &   0 & -a_1 a_2 J_3 \\
a_1 a_3 J_2 &   a_1 a_2 J_3 & 0 \\
\end{array}
\right).
\end{align}
In particular, evaluating $\mbox{Tr}\left( L^n\right)$, $\mbox{Tr}\left( R^n\right)$, $n$ being a positive integer, one can reproduce two constants of the motion found in the preceding section
\be
\mbox{Tr}\left( L^2\right)=-2 \mathcal{I}, \qquad  \mbox{Tr}\left( R^3\right)=6 \Omega.
\ee

The remaining integrals of motion follow by introducing the superpartners of the Lax matrices
\begin{align}\label{SP}
&
\delta_\epsilon L=W \epsilon,  && W={\rm i} \left(
\begin{array}{cccc}
0 & a_2 \theta_2 J_3-a_3 \theta_3 J_2  & -a_1 \theta_1 J_3+a_3 \theta_3 J_1 \\
-a_2 \theta_2 J_3+a_3 \theta_3 J_2 &   0 & a_1 \theta_1 J_2-a_2 \theta_2 J_1  \\
a_1 \theta_1 J_3-a_3 \theta_3 J_1& -a_1 \theta_1 J_2+a_2 \theta_2 J_1 & 0 \\
\end{array}
\right),
\nonumber\\[6pt]
&
\delta_\epsilon R=Z \epsilon, &&
Z=\left(
\begin{array}{cccc}
0 & a_1 J_1  & a_2 J_2 \\
-a_1 J_1  &   0 & a_3 J_3 \\
-a_2 J_2   &  -a_3 J_3  & 0 \\
\end{array}
\right),
\end{align}
and computing mixed traces
\be
\mbox{Tr}\left(R Z\right)=-2 Q, \qquad \mbox{Tr}\left(R^2 Z\right)=-{\rm i} \Lambda, \qquad \mbox{Tr}\left(LM-\frac 12 Z^2\right)=-2 \mathcal{H}.
\ee

As is well known, a Lax pair for the Euler top is not unique (see e.g. \cite{BBT}). It would be interesting to construct an alternative, which would allow one to reproduce all the conserved charged without invoking mixed traces.

\vspace{0.5cm}

\noindent
{\bf 4. The issue of integrability}\\

\noindent
When discussing the issue of integrability, it is customary to use constants of the motion to algebraically express some dynamical variables in terms of others. If a first integral is quadratic (or higher) in fermions, one faces the difficulty in resolving the equation as there does not exist a division by a Grassmann--odd number. Yet, because fermionic degrees of freedom for the case at hand obey first order ordinary differential equations, a general solution involves three Grassmann--odd constants of integration. Denoting them by $\alpha$, $\beta$ and $\gamma$ and taking into account $\alpha^2=\beta^2=\gamma^2=0$, one arrives at natural decompositions
\bea\label{comp}
&&
\theta_i=\alpha \varphi_{i1} + \beta \varphi_{i2}+\gamma \varphi_{i3} + {\rm i} \alpha \beta \gamma \varphi_{i4}, \qquad
J_i =J_{i 0} + {\rm i} \alpha\beta J_{ i 1}  +{\rm i} \alpha\gamma J_{ i 2}  +{\rm i} \beta\gamma J_{ i 3},
\eea
where components accompanying $\alpha$, $\beta$, and $\gamma$ are real {\it bosonic} functions of the temporal variable. Substituting (\ref{comp}) into the equations of motion (\ref{seom}) and analysing monomials in $\alpha$, $\beta$, $\gamma$, one turns (\ref{seom}) into a system of $24$ first order ordinary differential equations (see Appendix).

According to the Jacobi last multiplier method (see e.g. \cite{W}), a system of equations ${\dot x}_i=f_i (x)$, $i=1,\dots,n+1$, is integrable by quadratures if it possesses $n-1$ functionally independent first integrals and admits an integrating multiplier $\mu$ obeying
\be
\dot\mu+\mu \partial_i f_i=0.
\ee
If a vector field $f_i$ is divergence free, a system automatically admits an integrating multiplier $\mu=\mbox{const}$.

A brief glance at the equations in Appendix shows $\partial_i f_i=0$, which means that one needs $22$ first integrals in order to establish integrability. Substituting (\ref{comp}) into the conserved charges  $\mathcal{H}$, $\mathcal{I}$, $Q$, $\Lambda$, one obtains $15$ functionally independent first integrals (see Appendix). A closer examination of the component equations in Appendix reveals three more constants of the motion
\be\label{suppl}
\varphi_{11}^2+\varphi_{21}^2+\varphi_{31}^2, \qquad \varphi_{12}^2+\varphi_{22}^2+\varphi_{32}^2, \qquad \varphi_{13}^2+\varphi_{23}^2+\varphi_{33}^2,
\ee
of which only two prove to be functionally independent of the $15$ first integrals resulting from $\mathcal{I}$, $\mathcal{H}$, $Q$, $\Lambda$. Note that (\ref{suppl}) do not show up within the framework of the formalism in Sect. 2 because $\alpha^2=\beta^2=\gamma^2=0$. As $5$ first integrals are still missing, one is forced to conclude that the dynamical system (\ref{seom}) lacks the integrability property.

It is interesting to study whether the component form of Eqs. (\ref{seom}) given in Appendix admits integrable reductions. Setting $\gamma=0$ in (\ref{comp}), one eliminates $\varphi_{13}$, $\varphi_{14}$, $\varphi_{23}$, $\varphi_{24}$, $\varphi_{33}$, $\varphi_{34}$, $J_{12}$, $J_{13}$, $J_{22}$, $J_{23}$, $J_{32}$, $J_{33}$ from the consideration and obtains a system of $12$ equations of motion admitting $9$ first integrals. One integral is still missing for securing integrability in accord with the Jacobi last multiplier method.

Setting one more constant of integration in (\ref{comp}) to vanish, say $\beta=0$, one concentrates on a particular solution for which all fermions are proportional to one and the same Grassmann--odd number (a value of the conserved supercharge)
\be\label{vf}
\theta_i=\alpha\varphi_i, \qquad \alpha^2=0.
\ee
Because quadratic combinations of fermions now vanish, Eqs. (\ref{seom}) simplify to the form
\begin{align}\label{Rseom}
&
{\dot{J}}_1=\left(a_2^2-a_3^2 \right) J_2 J_3, && {\dot{J}}_2=-\left(a_1^2-a_3^2 \right) J_1 J_3,
\nonumber\\[2pt]
&
{\dot{J}}_3=\left(a_1^2-a_2^2 \right) J_1 J_2, && \dot\varphi_1=a_1 a_2 \varphi_2 J_3-a_1 a_3 \varphi_3 J_2,
\nonumber\\[2pt]
&
\dot\varphi_2=-a_1 a_2 \varphi_1 J_3+a_2 a_3 \varphi_3 J_1, && \dot\varphi_3=a_1 a_3 \varphi_1 J_2-a_2 a_3 \varphi_2 J_1,
\end{align}
which holds invariant under the transformation
\be\label{trphi}
J'_i=J_i, \qquad \varphi'_i=\varphi_i+a_{\hat i} J_{\hat i} \lambda,
\ee
where $\lambda$ is a Grassmann--even parameter, as well as under the rescaling $\varphi'_i=\nu \varphi_i$. The system (\ref{Rseom}) admits four functionally independent first integrals
\be\label{CONST}
J_i J_i, \qquad a_i^2 J_i^2, \qquad a_i \varphi_i J_i, \qquad  \varphi_i \varphi_i,
\ee
as well as the Jacobi last multiplier and, hence, is integrable by quadratures. Within the context of a rigid body dynamics, a similar model was discussed in \cite{AT}.

Because the dynamics of $J_i$  decouples in (\ref{Rseom}), the known solutions to (\ref{eom}) can be used to analyse the evolution of $\varphi_i$ over time.
In particular, focusing on a symmetric Euler top (\ref{sEt}), one gets
\be
\varphi_1(t)= \frac{a_1 c_1}{2 H}+\kappa \cos{\left(a_1 \sqrt{2H} t\right)}+\sigma\sin{\left(a_1 \sqrt{2H} t\right)},
\ee
where  $\kappa$, $\sigma$ are real constants of integration and $H=\frac 12 a_i^2 J_i^2 =\frac 12 \left(a_1^2 c_1^2+a_2^2 c_2^2 \right)$, while $\varphi_2$ and $\varphi_3$ follow from the formulae
\be
\varphi_2(t)=\frac{a_1 J_2 -a_1^2 J_1 J_2 \varphi_1+J_3 {\dot\varphi}_1}{a_1 a_2 c_2^2}, \qquad \varphi_3(t)=\frac{a_1 J_3 -a_1^2 J_1 J_3 \varphi_1-J_2 {\dot\varphi}_1}{a_1 a_2 c_2^2}.
\ee
This solution is further simplified for a spherical Euler top (\ref{sol1}) in which case the orbit is an intersection of a cone  and a sphere (two last constants of the motion in (\ref{CONST})). Note that the higher order fermionic invariants
(\ref{Om}) and (\ref{La}) vanish for the particular solution (\ref{vf}) as a consequence of $\alpha^2=0$.

For unequal moments of inertia $J_i$ are represented by elliptic functions and it proves easier to analyse Eqs. (\ref{Rseom}) numerically. For example, Fig. 1 and Fig. 2 display parametric plots of $J_i(t)$ and $\varphi_i (t)$ for $a_1=0.5$, $a_2=0.3$, $a_3=0.2$, $J_i(0)=1$, $\varphi_i(0)=1$, $t \in [0,500]$. Interestingly enough, even small deviations of the $J_i$--orbit from a circle on a sphere cause quite a significant distortion of the curve in the $\varphi_i$--subspace.

%======================= Fig.1===========================>
\begin{figure}[ht]
\begin{center}
\resizebox{0.28\textwidth}{!}{%
\includegraphics{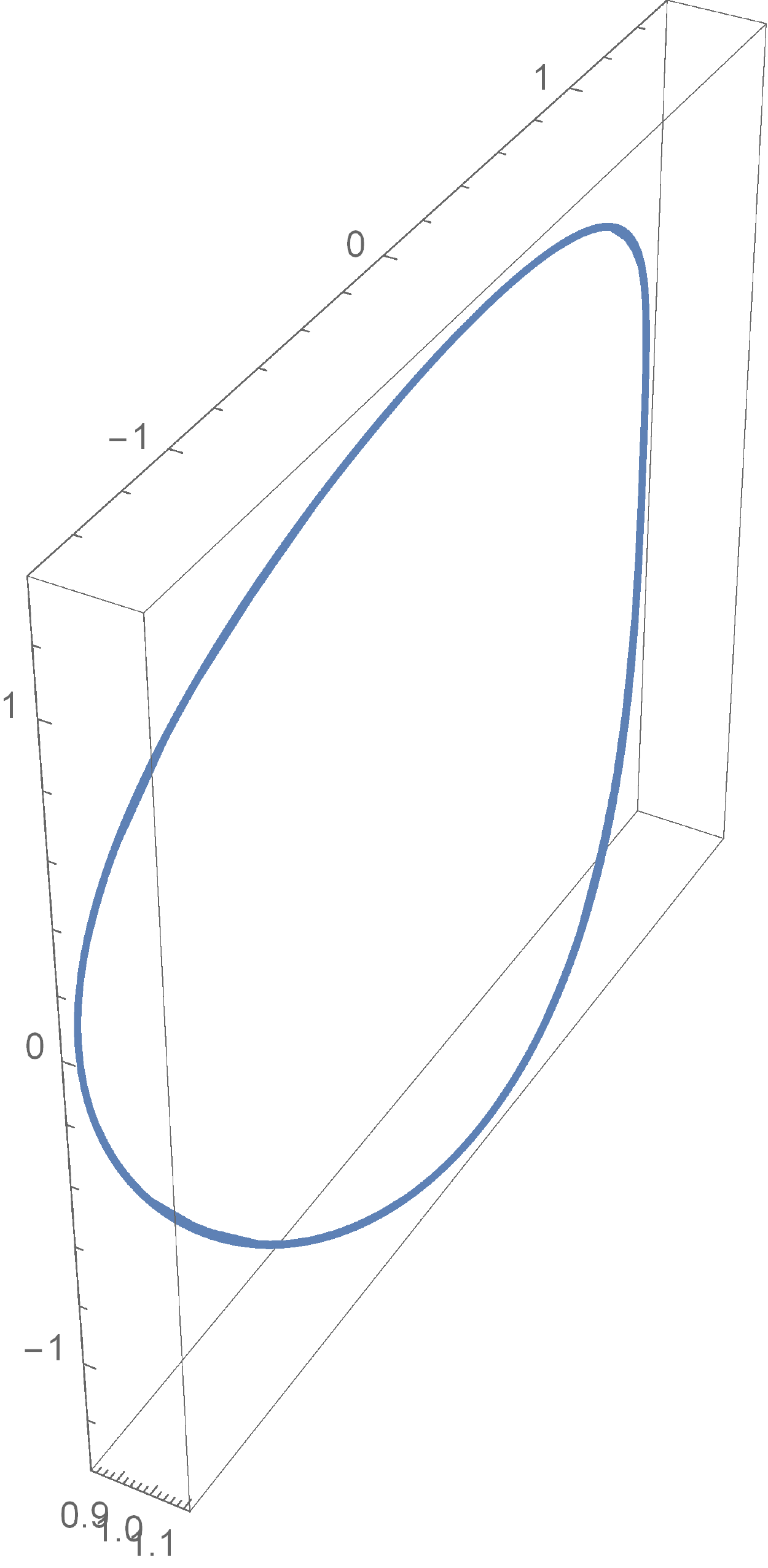}}\vskip-4mm
\caption{\small A parametric plot of $J_i(t)$ for $a_1=0.5$, $a_2=0.3$, $a_3=0.2$, $J_i(0)=1$, $\varphi_i (0)=1$, $t \in [0,500]$.}
\label{fig3}
\end{center}
\end{figure}

%======================= Fig.2===========================>
\begin{figure}[ht]
\begin{center}
\resizebox{0.40\textwidth}{!}{%
\includegraphics{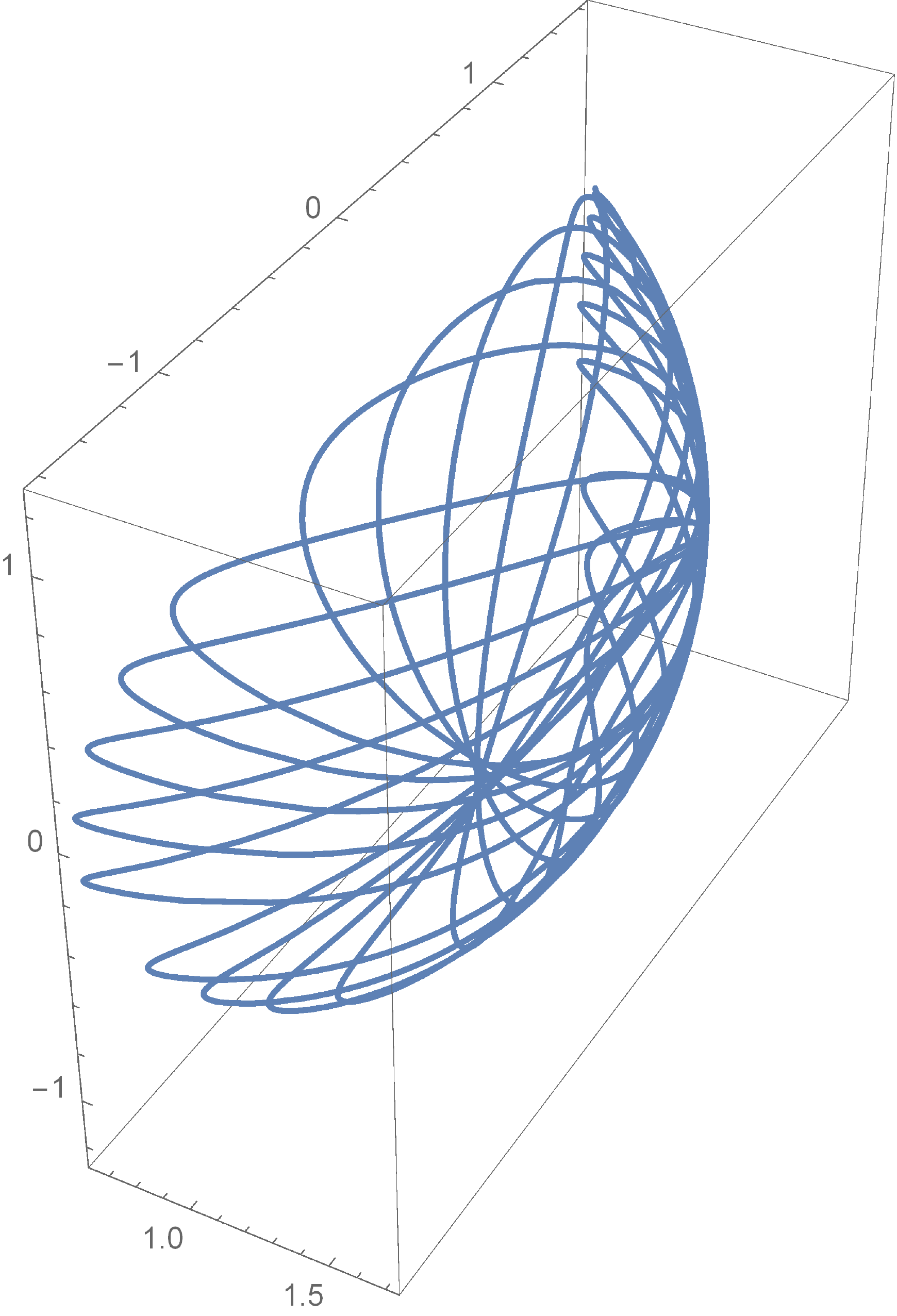}}\vskip-4mm
\caption{\small A parametric plot of $\varphi_i(t)$ for $a_1=0.5$, $a_2=0.3$, $a_3=0.2$, $J_i(0)=1$, $\varphi_i (0)=1$, $t \in [0,500]$.}
\label{fig3}
\end{center}
\end{figure}

As is well known, in some cases one can obtain a general solution to a differential equation by applying symmetry transformations to its particular solution. Implementing the supersymmetry transformation given in (\ref{comp1}) to a particular solution above, one gets the set of functions\footnote{Note that, given two Grassmann--odd numbers $\alpha$ and $\epsilon$, in general there does not exist a Grassmann--even number $\lambda$ such that $\epsilon=\lambda \alpha$. This follows from the fact that a product of two Grassmann--odd numbers is in general not zero. For this reason, the second term in $\varphi_i \alpha+a_{\hat i} J_{\hat i} \epsilon$ cannot be removed by redefining $\varphi_i$ in accord with the transformation (\ref{trphi}).}
\be\label{mod}
\theta'_i=\varphi_i \alpha+a_{\hat i} J_{\hat i} \epsilon, \qquad J'_i= J_i+{\rm i} \epsilon_{ijk} a_j \varphi_j J_k \alpha \epsilon,
\ee
where $\epsilon$ is a Grassmann--odd parameter and $J_i$, $\varphi_i$ are assumed to obey (\ref{Rseom}). Making use of the fact that $a_i J_i \varphi_i$ is a constant of the motion and the identity
\be
a_i \theta'_i J'_i=a_i \theta_i J_i+a_i^2 J_i^2 \epsilon,
\ee
one can verify that (\ref{mod}) does solve the full nonlinear system (\ref{seom}). Yet, the explicit calculation of $\Lambda$ in accord with (\ref{La}) gives zero. Thus, despite the fact that (\ref{mod}) involves two Grassmann--odd parameters $\alpha$ and $\epsilon$, it belongs to a class of solutions which only differ by a value of the conserved supersymmetry charge and have vanishing $\Lambda$. A solution deriving from $\alpha$--transformation in (\ref{comp}) belongs to the $\Lambda=0$ class as well.

Concluding this section, we list equilibrium points of the reduced system (\ref{Rseom})\footnote{Because $\varphi_i$ is defined up to a constant factor, one can set $a_i J_i \varphi_i=1$.}
\begin{align}\label{fp}
&
J_1=c_1, && J_{2,3}=0, && \varphi_1=\frac{1}{a_1 c_1}, && \varphi_{2,3}=0;
\nonumber\\[2pt]
&
J_2=c_2, && J_{1,3}=0, && \varphi_2=\frac{1}{a_2 c_2}, && \varphi_{1,3}=0;
\nonumber\\[2pt]
&
J_3=c_3, && J_{1,2}=0, && \varphi_3=\frac{1}{a_3 c_3}, && \varphi_{1,2}=0,
\end{align}
where $c_i$ are real constants. One can readily verify that the first and the last equilibria are stable against small perturbations, while the second option is not. It is worth mentioning that substituting the fixed points of the Euler top into the equations of motion for $\varphi_i$, one reveals that one component of $\varphi_i$ is constant, while two others obey one and the same harmonic oscillator equation.

\vspace{0.5cm}
\noindent
{\bf 5. A generalisation involving a three--dimensional real Lie algebra}\\

\noindent
The construction in Sect. 2 can be generalised to the case of an arbitrary three-dimensional real Lie algebra. Classification of such algebras was accomplished by Bianchi and the available options are displayed below in Table 1 (we follow a modern exposition in \cite{DNF}, in which $a$ designates an arbitrary real constant).
\begin{center}
Table 1. The Bianchi classification of three-dimensional real Lie algebras.
\end{center}
\begin{eqnarray*}
\footnotesize
\begin{array}{|l|r|r|r|r|r|r|r|r|c|}
\hline
  & \{J_1,J_2 \}  & \{J_1,J_3 \}   & \{J_2,J_3 \}  & \mbox{Casimir invariant}~ \mathcal{I} \\
  \hline
~ \mbox{type I} & 0 & 0 & 0 &  J_1, J_2, J_3 \\
\hline
~ \mbox{type II} & 0 & 0 & J_1 &  J_1 \\
\hline
\mbox{ type III} & J_2-J_3  & -J_2+J_3 & 0 & J_2+J_3\\
\hline
\mbox{ type IV} & J_2+J_3 & J_3 & 0 & \frac{J_2}{J_3}-\ln{J_3} \\
\hline
\mbox{ type V} & J_2 & J_3 & 0 & \frac{J_2}{J_3}\\
\hline
\mbox{ type VI} & a J_2-J_3 & -J_2+a J_3 & 0 & J_3^2 {\left(1+\frac{J_2}{J_3} \right)}^{1+a} {\left(1-\frac{J_2}{J_3} \right)}^{1-a} \\
\hline
~ \mbox{type $$VI$_0$} & 0 & J_2 & J_1& J_1^2-J_2^2  \\
\hline
 \mbox{ type VII} & a J_2+J_3 & -J_2+a J_3 & 0 & (J_2^2 + J_3^2) e^{-2 a \arctan{\frac{J2}{J3}}}  \\
\hline
\mbox{ type $$VII$_0$} & 0 & -J_2& J_1 & J_1^2+J_2^2  \\
\hline
\mbox{ type VIII} & -J_3& -J_2 & J_1 & J_1^2+J_2^2-J_3^2  \\
\hline
 \mbox{ type IX} & J_3 & -J_2 & J_1 & J_1^2+J_2^2+J_3^2  \\
\hline
\end{array}
\end{eqnarray*}

\vspace{0.2cm}

Let us consider a three-dimensional real Lie algebra with generators $J_i$, $i=1,2,3$, and structure constants $c_{ij}^k$. Identifying $J_i$ with bosonic degrees of freedom obeying the Poisson bracket
\be
\{J_i,J_j \}=c_{ij}^k J_k,
\ee
and
introducing real fermionic partners $\theta_i$ satisfying
\be
\{\theta_i,\theta_j \}=-{\rm i} \delta_{ij}, \qquad \{\theta_i,J_j \}=0,
\ee
one can construct the supersymmetry charge similar to (\ref{sc})
\be\label{gsc}
Q=a_i \theta_i J_i,
\ee
where $a_i$ are real constants. The latter gives rise to the Hamiltonian
\be\label{SHam1}
\qquad \{Q,Q \}=-2 {\rm i} \mathcal{H} \quad \rightarrow \quad
\mathcal{H}=\frac 12 a_i^2 J_i^2 +\frac{\rm i}{2} c_{ij}^k a_i \theta_i a_j \theta_j J_k,
\ee
which governs the dynamics.

By construction, the Casimir element in Table 1 provides an integral of motion of the dynamical system (\ref{SHam1}). One more invariant can be constructed by analogy with our consi\-deration in Sect. 2.
Taking into account the fermionic equations of motion
\be
{\dot\theta}_i=a_{\hat i} c_{\hat i j}^k a_j \theta_j J_k,
\ee
and the fact that the structure constants $c_{ij}^k$ are antisymmetric in the lower indices, one concludes that the cubic fermionic combination
\be\label{OMM}
\Omega=\frac{1}{3!} \epsilon_{ijk} a_i \theta_i a_j \theta_j a_k \theta_k
\ee
is conserved over time. Computing the bracket of (\ref{OMM}) with the supercharge, one obtains the invariant which is quadratic in fermions
\be
\{Q,\Omega \}=-\frac 12 \Lambda,  \qquad \Lambda={\rm i} \epsilon_{ijk} a_i \theta_i a_j \theta_j a_k^2 J_k,
\ee
and the identity involving $Q$, $\Lambda$, $\Omega$ and $\mathcal{H}$
\be
Q \Lambda=4 {\rm i} \mathcal{H} \Omega.
\ee
The latter implies that $\Omega$ is functionally dependent. Note that for the Bianchi type--II algebra one has an extra fermionic constant of the motion $\theta_1$.

\vspace{0.5cm}

\noindent
{\bf 6. Conclusion}\\

\noindent
To summarise, in this work a natural $\mathcal{N}=1$ supersymmetric extension of the Euler top was constructed, which introduces exactly one fermionic counterpart for each bosonic degree of freedom.
The equations of motion, their symmetries and integrals of motion were given. It was demonstrated that, although in general the system lacks the integrability property, it admits an interesting integrable reduction, which corresponds to a particular solution for which all fermions are proportional to one and the same Grassmann--odd number - a value of the conserved supercharge. A generalisation involving an arbitrary three--dimensional real Lie algebra was proposed.

Turning to possible further developments, it would be interesting to study supersymmetric extensions of the Euler top, which accommodate less than three fermions. This could resolve the issue of integrability \cite{AN}.
Our analysis in Sect. 4 shows that there are subtleties in treating integrals of motion which are quadratic in fermions or higher. It would be interesting to reconsider the issue of integrability of the supersymmetric extensions of the Calogero, Ruijsenaars--Schneider, and Toda systems proposed recently in \cite{AG2,AG3,KL}. In the latter regard, a similarity transformation in \cite{GLP} might be the key.

\vspace{0.5cm}

\noindent{\bf Acknowledgements}\\

\noindent
The author thanks Armen Nersessian for helpful discussions.
This work is supported by the Russian Foundation for Basic Research, grant No 20-52-12003.

\vspace{0.5cm}

\noindent
{\bf Appendix}\\

\noindent
In this Appendix, we gather the differential equations,  which follow from (\ref{seom}) after introducing the decompositions (\ref{comp})
\bea
&&
{\dot J}_{10}=\left(a_2^2-a_3^2 \right) J_{20} J_{30},
\nonumber\\[2pt]
&&
{\dot J}_{11}=\left(a_2^2-a_3^2 \right) \left(J_{20} J_{31}+J_{30} J_{21} \right)-a_1 a_2 J_{20} \left(\varphi_{11} \varphi_{22}-\varphi_{12} \varphi_{21} \right)
-a_1 a_3 J_{30} \left(\varphi_{11} \varphi_{32}-\varphi_{12} \varphi_{31} \right),
\nonumber\\[2pt]
&&
{\dot J}_{12}=\left(a_2^2-a_3^2 \right) \left(J_{20} J_{32}+J_{30} J_{22} \right)-a_1 a_2 J_{20} \left(\varphi_{11} \varphi_{23}-\varphi_{13} \varphi_{21} \right)
-a_1 a_3 J_{30} \left(\varphi_{11} \varphi_{33}-\varphi_{13} \varphi_{31} \right),
\nonumber\\[2pt]
&&
{\dot J}_{13}=\left(a_2^2-a_3^2 \right) \left(J_{20} J_{33}+J_{30} J_{23} \right)-a_1 a_2 J_{20} \left(\varphi_{12} \varphi_{23}-\varphi_{13} \varphi_{22} \right)
-a_1 a_3 J_{30} \left(\varphi_{12} \varphi_{33}-\varphi_{13} \varphi_{32} \right);
\nonumber\\[8pt]
&&
{\dot J}_{20}=-\left(a_1^2-a_3^2 \right) J_{10} J_{30},
\nonumber\\[2pt]
&&
{\dot J}_{21}=-\left(a_1^2-a_3^2 \right) \left(J_{10} J_{31}+J_{30} J_{11} \right)+a_1 a_2 J_{10} \left(\varphi_{11} \varphi_{22}-\varphi_{12} \varphi_{21} \right)
-a_2 a_3 J_{30} \left(\varphi_{21} \varphi_{32}-\varphi_{22} \varphi_{31} \right),
\nonumber\\[2pt]
&&
{\dot J}_{22}=-\left(a_1^2-a_3^2 \right) \left(J_{10} J_{32}+J_{30} J_{12} \right)+a_1 a_2 J_{10} \left(\varphi_{11} \varphi_{23}-\varphi_{13} \varphi_{21} \right)
-a_2 a_3 J_{30} \left(\varphi_{21} \varphi_{33}-\varphi_{23} \varphi_{31} \right),
\nonumber\\[2pt]
&&
{\dot J}_{23}=-\left(a_1^2-a_3^2 \right) \left(J_{10} J_{33}+J_{30} J_{13} \right)+a_1 a_2 J_{10} \left(\varphi_{12} \varphi_{23}-\varphi_{13} \varphi_{22} \right)
-a_2 a_3 J_{30} \left(\varphi_{22} \varphi_{33}-\varphi_{23} \varphi_{32} \right);
\nonumber\\[8pt]
&&
{\dot J}_{30}=\left(a_1^2-a_2^2 \right) J_{10} J_{20},
\nonumber\\[2pt]
&&
{\dot J}_{31}=\left(a_1^2-a_2^2 \right) \left(J_{10} J_{21}+J_{20} J_{11} \right)+a_1 a_3 J_{10} \left(\varphi_{11} \varphi_{32}-\varphi_{12} \varphi_{31} \right)
+a_2 a_3 J_{20} \left(\varphi_{21} \varphi_{32}-\varphi_{22} \varphi_{31} \right),
\nonumber\\[2pt]
&&
{\dot J}_{32}=\left(a_1^2-a_2^2 \right) \left(J_{10} J_{22}+J_{20} J_{12} \right)+a_1 a_3 J_{10} \left(\varphi_{11} \varphi_{33}-\varphi_{13} \varphi_{31} \right)
+a_2 a_3 J_{20} \left(\varphi_{21} \varphi_{33}-\varphi_{23} \varphi_{31} \right),
\nonumber\\[2pt]
&&
{\dot J}_{33}=\left(a_1^2-a_2^2 \right) \left(J_{10} J_{23}+J_{20} J_{13} \right)+a_1 a_3 J_{10} \left(\varphi_{12} \varphi_{33}-\varphi_{13} \varphi_{32} \right)
+a_2 a_3 J_{20} \left(\varphi_{22} \varphi_{33}-\varphi_{23} \varphi_{32} \right);
\nonumber\\[2pt]
&&
{\dot\varphi}_{11}=a_1 a_2 \varphi_{21} J_{30}-a_1 a_3 \varphi_{31} J_{20}, \qquad {\dot\varphi}_{12}=a_1 a_2 \varphi_{22} J_{30}-a_1 a_3 \varphi_{32} J_{20},
\nonumber\\[2pt]
&&
{\dot\varphi}_{13}=a_1 a_2 \varphi_{23} J_{30}-a_1 a_3 \varphi_{33} J_{20},
\qquad
{\dot\varphi}_{14}=a_1 a_2\left( \varphi_{21} J_{33}-\varphi_{22} J_{32}+ \varphi_{23} J_{31}+\varphi_{24} J_{30}\right)
\nonumber\\[2pt]
&&
\qquad \qquad \qquad \qquad \qquad \qquad \qquad \qquad \quad \quad
-a_1 a_3\left( \varphi_{31} J_{23}-\varphi_{32} J_{22}+ \varphi_{33} J_{21}+\varphi_{34} J_{20}\right);
\nonumber
\eea
\bea
&&
{\dot\varphi}_{21}=-a_1 a_2 \varphi_{11} J_{30}+a_2 a_3 \varphi_{31} J_{10}, \qquad {\dot\varphi}_{22}=-a_1 a_2 \varphi_{12} J_{30}+a_2 a_3 \varphi_{32} J_{10},
\nonumber\\[2pt]
&&
{\dot\varphi}_{23}=-a_1 a_2 \varphi_{13} J_{30}+a_2 a_3 \varphi_{33} J_{10},
\qquad
{\dot\varphi}_{24}=-a_1 a_2\left( \varphi_{11} J_{33}-\varphi_{12} J_{32}+ \varphi_{13} J_{31}+\varphi_{14} J_{30}\right)
\nonumber\\[2pt]
&&
\qquad \qquad \qquad \qquad \qquad \qquad \qquad \qquad \qquad \quad
+a_2 a_3\left( \varphi_{31} J_{13}-\varphi_{32} J_{12}+ \varphi_{33} J_{11}+\varphi_{34} J_{10}\right);
\nonumber\\[2pt]
&&
{\dot\varphi}_{31}=a_1 a_3 \varphi_{11} J_{20}-a_2 a_3 \varphi_{21} J_{10}, \qquad {\dot\varphi}_{32}=a_1 a_3 \varphi_{12} J_{20}-a_2 a_3 \varphi_{22} J_{10},
\nonumber\\[2pt]
&&
{\dot\varphi}_{33}=a_1 a_3 \varphi_{13} J_{20}-a_2 a_3 \varphi_{23} J_{10},
\qquad
{\dot\varphi}_{34}=a_1 a_3\left( \varphi_{11} J_{23}-\varphi_{12} J_{22}+ \varphi_{13} J_{21}+\varphi_{14} J_{20}\right)
\nonumber\\[2pt]
&&
\qquad \qquad \qquad \qquad \qquad \qquad \qquad \qquad \quad \quad
-a_2 a_3\left( \varphi_{21} J_{13}-\varphi_{22} J_{12}+ \varphi_{23} J_{11}+\varphi_{24} J_{10}\right).
\nonumber
\eea

Similarly, the integrals of motion $\mathcal{I}$, $\mathcal{H}$, $Q$, and $\Lambda$ reduce to
\bea
&&
\mathcal{I}_1=J_{10}^2+J_{20}^2+J_{30}^2, \qquad \mathcal{I}_2=J_{10} J_{11}+J_{20} J_{21}+J_{30} J_{31}, \qquad
\mathcal{I}_3=J_{10} J_{12}+J_{20} J_{22}+J_{30} J_{32},
\nonumber\\[2pt]
&&
\mathcal{I}_4=J_{10} J_{13}+J_{20} J_{23}+J_{30} J_{33};
\nonumber\\[8pt]
&&
\mathcal{H}_1=a_1^2 J_{10}^2+a_2^2 J_{20}^2+a_3^2 J_{30}^2, \qquad
\mathcal{H}_2=a_1^2 J_{10} J_{11}+a_2^2 J_{20} J_{21}+a_3^2 J_{30} J_{31}
\nonumber\\[2pt]
&&
+a_1 a_2 J_{30} \left( \varphi_{11} \varphi_{22}-\varphi_{12} \varphi_{21} \right)
-a_1 a_3 J_{20} \left( \varphi_{11} \varphi_{32}-\varphi_{12} \varphi_{31} \right)
+a_2 a_3 J_{10} \left( \varphi_{21} \varphi_{32}-\varphi_{22} \varphi_{31} \right),
\nonumber\\[2pt]
&&
\mathcal{H}_3=a_1^2 J_{10} J_{12}+a_2^2 J_{20} J_{22}+a_3^2 J_{30} J_{32}
+a_1 a_2 J_{30} \left( \varphi_{11} \varphi_{23}-\varphi_{13} \varphi_{21} \right)-
a_1 a_3 J_{20} \left( \varphi_{11} \varphi_{33}-\varphi_{13} \varphi_{31} \right)
\nonumber\\[2pt]
&&
+a_2 a_3 J_{10} \left( \varphi_{21} \varphi_{33}-\varphi_{23} \varphi_{31} \right), \qquad
\mathcal{H}_4=a_1^2 J_{10} J_{13}+a_2^2 J_{20} J_{23}+a_3^2 J_{30} J_{33}
\nonumber\\[2pt]
&&
+a_1 a_2 J_{30} \left( \varphi_{12} \varphi_{23}-\varphi_{13} \varphi_{22} \right)-
a_1 a_3 J_{20} \left( \varphi_{12} \varphi_{33}-\varphi_{13} \varphi_{32} \right)
+a_2 a_3 J_{10} \left( \varphi_{22} \varphi_{33}-\varphi_{23} \varphi_{32} \right); \qquad
\nonumber\\[8pt]
&&
Q_1=a_1 \varphi_{11} J_{10}+a_2 \varphi_{21} J_{20}+a_3 \varphi_{31} J_{30}, \qquad Q_2=a_1 \varphi_{12} J_{10}+a_2 \varphi_{22} J_{20}+a_3 \varphi_{32} J_{30},
\nonumber\\[8pt]
&&
Q_3=a_1 \varphi_{13} J_{10}+a_2 \varphi_{23} J_{20}+a_3 \varphi_{33} J_{30}, \qquad Q_4=a_1 \left(\varphi_{11} J_{13}-\varphi_{12} J_{12}+\varphi_{13} J_{11}+\varphi_{14} J_{10} \right)
\nonumber\\[8pt]
&&
+a_2 \left(\varphi_{21} J_{23}-\varphi_{22} J_{22}+\varphi_{23} J_{21}+\varphi_{24} J_{20} \right)+a_3 \left(\varphi_{31} J_{33}-\varphi_{32} J_{32}+\varphi_{33} J_{31}+\varphi_{34} J_{30} \right);
\nonumber\\[8pt]
&&
\Lambda_1=a_1 J_{10}  \left( \varphi_{21} \varphi_{32}-\varphi_{22} \varphi_{31} \right)-a_2 J_{20}  \left( \varphi_{11} \varphi_{32}-\varphi_{12} \varphi_{31} \right)+
a_3 J_{30}  \left( \varphi_{11} \varphi_{22}-\varphi_{12} \varphi_{21} \right),
\nonumber\\[8pt]
&&
\Lambda_2=a_1 J_{10}  \left( \varphi_{21} \varphi_{33}-\varphi_{23} \varphi_{31} \right)-a_2 J_{20}  \left( \varphi_{11} \varphi_{33}-\varphi_{13} \varphi_{31} \right)+
a_3 J_{30}  \left( \varphi_{11} \varphi_{23}-\varphi_{13} \varphi_{21} \right),
\nonumber\\[8pt]
&&
\Lambda_3=a_1 J_{10}  \left( \varphi_{22} \varphi_{33}-\varphi_{23} \varphi_{32} \right)-a_2 J_{20}  \left( \varphi_{12} \varphi_{33}-\varphi_{13} \varphi_{32} \right)+
a_3 J_{30}  \left( \varphi_{12} \varphi_{23}-\varphi_{13} \varphi_{22} \right).
\nonumber
\eea

\end{document}